\documentclass[3p,shortbib,onecolumn]{elsarticle}

\usepackage{amsmath,amssymb}

\usepackage{graphicx}

\usepackage[usenames]{color}

\newcommand{\be}{\begin{equation}}
\newcommand{\ee}{\end{equation}}
\newcommand{\bea}{\begin{eqnarray}}
\newcommand{\eea}{\end{eqnarray}}
\newcommand{\ud}{\mathrm{d}}
\newcommand{\bi}{\begin{itemize}}
\newcommand{\ei}{\end{itemize}}

\makeatletter
\def\ps@pprintTitle{%
 \let\@oddhead\@empty
 \let\@evenhead\@empty
 \def\@oddfoot{}%
 \let\@evenfoot\@oddfoot}
\makeatother

\begin{document}

\begin{frontmatter}

\title{Screw-symmetric gravitational waves: a double copy of the vortex}
\author{A.~Ilderton}
\ead{anton.ilderton@plymouth.ac.uk}
\address{Centre for Mathematical Sciences, University of Plymouth, PL4 8AA, UK}

\begin{abstract}
Plane gravitational waves can admit a sixth `screw' isometry beyond the usual five. The same is true of plane electromagnetic waves. 
From the point of view of integrable systems, a sixth isometry would appear to over-constrain particle dynamics in such waves; we show here, though, that no effect of the sixth isometry is independent of those from the usual five. Many properties of particle dynamics in a screw-symmetric gravitational wave are also seen in a (non-plane-wave) electromagnetic vortex; we make this connection explicit, showing that the screw-symmetric gravitational  wave is the classical double copy of the vortex.
\end{abstract}


\end{frontmatter}
%\clearpage
%\tableofcontents
%%%%%%%%%%%
\section{Introduction}
%%%%%%%%%%%
%
%
The plane wave approximation provides a simplified setting in which to investigate signatures of gravitational waves~\cite{Abbott:2017oio,TheLIGOScientific:2017qsa}, such as the velocity memory effect~\cite{Ehlers:1962zz,Zhang:2017rno,Zhang:2018srn}, in which particles initially at rest acquire a constant, nonzero velocity after the wave has passed over them. The same effect is seen in electromagnetic plane waves, see e.g.~\cite{Dinu:2012tj}, and~\cite{Zhang:2017geq} for historical references, with connections to the infra-red in both cases~\cite{Ilderton:2012qe,Zhang:2017geq}.  The possibility of mapping gravitational observables onto a simpler gauge theory setting~\cite{Goldberger:2016iau,Goldberger:2017frp} provides one motivation for studying the ``classical double copy''~\cite{Monteiro:2014cda}, that is the mapping of classical solutions of Einstein's equations to classical solutions of Yang-Mills' equations.  This is part of a larger program on colour-kinematic duality, or double copy, a precise conjecture about how scattering amplitudes in gravity can be obtained from those in gauge theory by replacing colour structure with kinematic structure~\cite{Bern:2008qj,Bern:2010ue,Bern:2010yg}. The double copy conjecture has been proven at tree level, and there are an increasing number of nontrivial examples at loop level, see~\cite{Carrasco:2015iwa} for a review. In this context we note that plane waves provide a testing ground for extending the double copy programme to curved backgrounds~\cite{Adamo:2017nia}.

There are noticeable similarities between particle motion in an electromagnetic vortex~\cite{BB-Vortex}, which is not a plane wave, and in certain circularly polarised gravitational waves. The latter have been investigated as models of the waves emitted in various astrophysical phenomena~\cite{Kamionkowski:2015yta,Abbott:2017oio,Abbott:2017mnu}. Such waves can show an enlarged symmetry group containing an additional `screw isometry'~\cite{Sippel:1986if,Zhang:2017rno,Zhang:2018srn} beyond the five common to all plane waves. Our focus here is on the role played by this (and other) additional symmetries in charge motion, and our goal is to tie this to related results in integrable systems, to dynamics in electromagnetic vortices, and to the classical double copy.

This paper is organised as follows. In Sect.~\ref{SECT:SUPER} we review the isometries of, and particle motion in, plane gravitational and electromagnetic waves. From the point of view of integrable systems these are rather special `superintegrable' systems. In Sect.~\ref{SECT:SCREW} we consider the screw isometry, which would seem to imply the existence of one conserved quantity too many. We resolve this, showing explicitly that the implied integral of motion is not independent of the other five. In Sect.~\ref{SECT:COPY} we compare charge motion in the screw-symmetric wave with that in an electromagnetic vortex~\cite{BB-Vortex}, finding many similarities. We make the connection concrete by observing that the screw-symmetric wave is the classical double copy of the vortex. We discuss related cases and conclude in Sect.~\ref{SECT:CONCS}.
%
%%%%%%%%
\section{Isometries and (super)-integrable motion in plane waves}\label{SECT:SUPER}
%%%%%%%%%%%%%%%%%%%%%%%%%
\subsection{Gravitational plane waves} 
%%%%%%%%%%%%%%%%%%%%%%%%%%%%%%%
In order to make symmetries manifest we begin in Baldwin-Jeffery-Rosen (BJR) coordinates $\{u,v,x^j\}$, where the plane wave metric has the form~\cite{Baldwin95,Rosen}
\be\label{PV-ER}
	g_{\mu\nu}\ud x^\mu \ud x^\nu =  \ud u \ud v - \gamma_{ij}(u) \ud x^i \ud x^j \;, \qquad j\in \{1,2\} \;.
\ee
These coordinates are not global, and the $\gamma_{ij}$ are constrained by the vacuum equations, but this will not affect our arguments. We will switch to globally defined coordinates later (see e.g.~\cite{Zhang:2017geq,Adamo:2017nia} for recent discussions and further references). The five Killing vectors of the metric are
\be\label{KILLING-ER}
	\bigg\{\frac{\partial}{\partial x^i} \;, \;  2x^i \frac{\partial}{\partial v} + G^{ij}(u)\frac{\partial}{\partial x^j}  \;, \;  \frac{\partial}{\partial v} \bigg\} \;, \quad\text{where} \quad  G^{ij}(u) = \int\limits^{u}\!\ud s\, \gamma^{ij}(s) \;,
\ee
corresponding to invariance under the Carroll group~\cite{Carroll} with broken rotations~\cite{Duval:2017els}. Now consider a test particle in this background. Each Killing vector implies the existence of a conserved quantity in the particle motion. To analyse this we use the Hamiltonian formalism, which requires gauging the reparameterisation invariance of the particle action as usual, and we take $u$ as time~\cite{Heinzl:2000ht}.  The action and Hamiltonian are
\be\label{H-GR}
	S = - m\int\sqrt{g_{\mu\nu}\ud x^\mu \ud x^\nu } \quad \longrightarrow\quad  H(u) = \frac{\gamma^{ij}(u)p_{i}p_j+ m^2}{4p_v} \;,
\ee
where $\{p_v,p_j\}$ are the respective conjugate momenta to $\{v,x^j\}$. The five conserved quantities corresponding to the five Killing vectors above are 
\be\label{GV-P}
	\{ Q_1, \cdots, Q_5 \} := \big\{p_j  \;, \;  2x^i p_v +G^{ij}(u) p_j \;, \;  p_v \big\} \;. 
\ee
The conservation of these five is enough to determine all momenta and $x^j$ algebraically, after which Hamilton's equation for $v$ may be integrated directly. The question we want to address is, how many conserved quantities can there be? To answer this we need some general results on integrable systems.

An autonomous Hamiltonian system with $2n$-dimensional phase space is (polynomially) superintegrable if there exist $N>n$ independent phase space functions $Q_j$ (polynomial in the momenta) which Poisson commute with the Hamiltonian (are conserved), and such that $n$ of them are in involution, $\{Q_i,Q_j\}=0\, \forall\, i,j\in\{1\ldots n\}$. Systems with $N=2n-1$, the maximum possible number, are called maximally superintegrable~\cite{Wojciechowski:1983,Miller:2013}. While $2n-1$ conserved functions always exist {\it locally}~\cite{Courant}, it is very rare to find systems in which they are globally defined polynomials in the momenta~\cite{Miller:2013}. Superintegrable systems have many appealing properties; the classical equations of motion can admit an algebraic solution, and there is a conjecture that all corresponding quantum systems are exactly solvable~\cite{Tempesta:2001}.  

A test particle in a gravitational plane wave is a superintegrable system; to show this, and noting that the Hamiltonian is time-dependent, we follow the standard method of converting to an autonomous system\footnote{Alternatively, taking $v$ to be time, rather than $u$, gives an autonomous system. However, to make connection to other cases it is more convenient if the wave depends on the choice of time. The same is often true in QED calculations~\cite{Neville:1971uc,Bakker:2013cea}.}; we expand phase space to eight dimensions by promoting $u$ to a coordinate with conjugate momentum $p_u$, and use a new Hamiltonian $K= H - p_u$, for a review see~\cite{Struck}. Writing a dash for a derivative with respect to a new time (which appears nowhere explicitly), the time-derivative of any quantity $Q$ is
\be\label{tids-deriv-definition}
	Q' = \{K,Q\}_* \quad \text{where} \quad \{A,B\}_* = \frac{\partial A}{\partial x^\mu}\frac{\partial B}{\partial p_\mu} - \frac{\partial B}{\partial x^\mu}\frac{\partial A}{\partial p_\mu} \;. 
\ee 
In particular, we have as usual 	$u' = -\partial K/\partial p_u= 1$. Now, in general there is no way to know \textit{a priori} if a given system is (super)integrable. To derive the conserved quantities one can simply make an ansatz for $Q$ (e.g.~that it is quadratic in momenta) and impose (\ref{tids-deriv-definition}); this yields a series of algebraic and differential equations determining the form of $Q$, see~\cite{Miller:2013,Marchesiello:2015,Heinzl:2017blq} for examples and references.  In our plane wave case, this procedure yields $Q_1\ldots Q_5$ as in (\ref{GV-P}), along with two further conserved quantities; $Q_6 = p_u p_v - p_v H(u)$, which is just the mass-shell condition, and $Q_7$, given by
\be
	Q_7 = 4p_v^2 v - m^2 u - G^{ij}(u)p_i p_j \;. 
\ee
These seven are functionally independent\footnote{Defining $\mathcal{F}=\{Q_1,\ldots Q_N\}$ and following~\cite{Miller:2013}, the $N$ quantities $Q_j$ are functionally independent if the $N\times 8$ matrix $\mathcal{M}$ has rank $N$, where
\be
	\mathcal{M}_{l\mu} := \bigg( \frac{\partial \mathcal{F}_l}{\partial x^\mu} , \frac{\partial \mathcal{F}_l}{\partial p_\mu} \bigg)  \quad \text{(no sum).}
\ee
}. Thus we have the maximum number of seven independent conserved quantities, polynomial in the momenta. The system is maximally polynomially superintegrable. The solution of the equations of motion proceeds algebraically from here: the three momenta are conserved, $Q_4$ and $Q_5$ then determine $\{x^1,x^2\}$ as functions of time $u$, while $Q_7$ determines $v$.

%%%%%%%%
\subsection{Electromagnetic plane waves}
%%%%%%%%
Let us compare with electromagnetic plane waves. We work in lightfront coordinates $\{u,v,x^1,x^2\}$, the metric being (\ref{PV-ER}) with $\gamma_{ij}(u)\to \delta_{ij}$. In order to make the connections with the gravitational case clear we represent an arbitrary electromagnetic plane wave $F_{\mu\nu}\equiv F_{\mu\nu}(u)$ using the two-component `BJR' potential
\be\label{A-GEN}
	A(x) = A_j(u) \ud x^j \;, \qquad j\in \{1,2\} \;.
\ee
The particle action and, again taking $u$ as time, (reparameterisation-)gauge-fixed Hamiltonian are now
\be\label{H-U1}
	S = - \int\!\ud\tau\, m\sqrt{{\dot x}.{\dot x}} + \dot{x}.A(x) \quad \longrightarrow\quad H(u) =  \frac{(p_j - A_j(u))^2 +m^2 }{4p_v} \;.
\ee
An arbitrary electromagnetic plane wave has five isometries, $\mathcal{L}_\xi F_{\mu\nu}=0$, for
\be\label{fem}
	\xi \in \bigg\{ \frac{\partial}{\partial x^j}  \;, 2x^j \frac{\partial}{\partial v} + u \frac{\partial}{\partial x^j} \;,  \frac{\partial}{\partial v} \bigg\} \;, 
\ee
corresponding to invariance under three translations and two null rotations respectively. These are of course in direct analogy to (\ref{KILLING-ER}) and again span the Carroll group with broken rotations.  Because these are Poincar\'e transformations they imply the existence of five conserved quantities; for $\xi \equiv \xi^\mu\partial_\mu$ Poincar\'e we have~\cite{Heinzl:2017blq}
\be\begin{split}\label{Q-xi-lambda}
	\mathcal{L}_\xi F_{\mu\nu} = 0  &\implies Q \equiv \xi(x).p - \Lambda(x)= \text{ constant, where } \mathcal{L}_\xi A_\mu = \partial_\mu \Lambda \;.
\end{split}
\ee
(The functions $\Lambda$ appear because the potential need only be symmetric up to U(1) gauge transformations.) The five conserved quantities following from the Poincar\'e symmetries of the plane wave (\ref{fem}) are
\be\label{U1-P}
	\{Q_1, \cdots, Q_5\} = \bigg\{ p_j  \;, \; 2 x^j p_v + u p_j - G_j(u) \;, \;  p_v\bigg\} \quad \text{for} \quad G_j(u) = \int\limits^{u}\!\ud s\, A_j(s) \;,
\ee
in which the integrals are gauge terms $\Lambda$ as in~(\ref{Q-xi-lambda}). These are again in analogy to (\ref{GV-P}). There are two further conserved quantities on expanded phase space; $Q_6$ is as above but with the Hamiltonian (\ref{H-GR}) replaced by (\ref{H-U1}), while
\be
	Q_7 = 4 p_v^2 v -(p_i p_i+m^2) u + 2p_j G_j(u) - \int\limits^u\!\ud s\, A_i(s) A_i(s) \;.
\ee
Thus we have again the maximal number of conserved quantities, polynomial in the momenta, and again the system is maximally polynomially superintegrable.

%%%%%%%%%%%%%%%%%%%%%%%%%%%%%%%%%%%%
\section{Screw symmetry}\label{SECT:SCREW}
%%%%%%%%%%%%%%%%%%%%%%%%%%%%%%%%%%%%%
%
In both the gravitational and electromagnetic cases the conserved quantities following from invariance under the Carroll transformations are sufficient to determine the particle orbit. These quantities are `universal', i.e.~they have the same form for any given plane wave profile and, together with those from the extended phase space, they exhaust the list of possible independent conserved quantities in the motion. Nevertheless, both gravitational and electromagnetic plane waves can have additional isometries~\cite{Sippel:1986if,Shore:2017dqx,Zhang:2017rno,Zhang:2018srn}. For example, both can admit the screw isometry generated by the sum of a translation and a rotation; we consider the consequences of this extra symmetry here.

For the electromagnetic case the screw-symmetric gauge field is $A_1 = a_0 m\cos \omega u$ and $A_2 = a_0 m\sin \omega u$. It is symmetric under the transformation
\be\label{screw1}
	\xi = \frac{\partial}{\partial u} + \omega \bigg(x^1 \frac{\partial}{\partial x^2} - x^2 \frac{\partial}{\partial x^1}\bigg) \;.
\ee
Because this is a Poincar\'e transformation it implies the conservation of
\be\label{screw2}
	Q_8 = \xi\cdot p = H(u) + \omega(x p_2-y p_1) \;,
\ee
in the particle motion, as may be verified using (\ref{H-U1}). However, the existence of an eighth conserved function on (extended) phase space contradicts the general results above. It follows that $Q_8$ cannot be independent of the others. This may be confirmed explicitly; for $\{Q_1\ldots Q_5\}$ the five universal quantities in (\ref{U1-P}), we have 
\be
	Q_8 = \frac{a_0^2+m^2+Q_1^2+Q_2^2+ 2 \omega(Q_2 Q_3-Q_1Q_4)}{4 Q_5} \;. 
\ee
Hence, because $Q_1 \ldots Q_5$ are conserved, $Q_8$ is automatically conserved. In the sense that motion is determined entirely by the universal symmetries, the screw isometry of the wave, or indeed any additional symmetry of $F_{\mu\nu}$, is `redundant'.  

In order to consider the gravitational case in detail we switch to the globally defined Brinkmann coordinates~\cite{Brinkmann:1925fr} $(U,V,X^j)$ with
\be\label{BRINK}
	\ud s^2 = \ud U \ud V - \ud X^j \ud X^j - H_{ij}(U)X^i X^j \ud U^2 \;,
\ee
and where the matrix $H(U)$ is traceless. Though explicit invariance of the metric under translations in $x^j$ and null rotations is now lost, these are mapped to four corresponding symmetries in the new coordinates. These can in general only be given implicitly as
\be\label{xi-4-grav}
	\xi = f_j(U) \frac{\partial}{\partial X^j} + 2 X^j {\dot f}_j(U) \frac{\partial}{\partial V} \quad \text{where} \quad \ddot{f}_i (U)= H_{ij}(U) f_j(U) \;.
\ee
There are four independent solutions to the equation on the right, each yielding a conserved quantity $\xi^\mu p_\mu$ in the particle motion. $\partial/\partial V$ is clearly still a killing vector, yielding a fifth conserved quantity. These are the universal five in Brinkmann coordinates. For the particular choice of a circularly polarised, monochromatic wave,
\be\label{H-VAL}
	H_{ij} = H_0\begin{pmatrix}
		\cos \omega U  & \sin \omega U \\
		\sin \omega U & - \cos \omega U
		\end{pmatrix} \;,
\ee
it is easily verified that
\be\label{SCREW}
	\xi = \frac{\partial}{\partial U} + \frac{\omega}{2} \bigg(X^1 \frac{\partial}{\partial X^2} - X^2 \frac{\partial}{\partial X^1}\bigg) \;,
\ee
which generates the screw transformation, is an additional Killing vector, implying an additional conserved quantity. As for the electromagnetic case, this cannot be independent of the other five. We will show this explicitly below. This is \textit{not} to say that the screw symmetry has no effect; it reflects the fact that the field has certain properties, and this implies certain signatures in particle motion. However, all properties of the motion are determined by the universal five symmetries.  The fact that the wave is so symmetric still makes it a very interesting case to study, and in the remainder of this paper we continue to investigate its properties and its connection to \textit{non}-plane wave electromagnetic fields.

%%%%%%%%%%%%%%%%%%%%%%%%%%%%%%
\section{Electromagnetic vortices and the double copy}\label{SECT:COPY}
%%%%%%%%%%%%%%%%%%%%%%%%%%%%%%

%%%%%%%%%%%%%%%%
\subsection{The double copy}
%%%%%%%%%%%%%%%%
%
Again using coordinates $g_{\mu\nu}\ud x^\mu \ud x^\nu = \ud U\ud V - \ud X^j \ud X^j$, the electromagnetic vortex~\cite{BB-Vortex} is a solution of Maxwell's equations in vacuum with electric and magnetic fields oscillating in $U$ and growing linearly with $|{\bf X}|$; it is not a plane wave. The vortex has only \textit{two} Poincar\'e symmetries~\cite{Heinzl:2017blq}, $\partial/\partial V$ and the screw symmetry~(\ref{SCREW}). Particle motion in the vortex is nevertheless exactly solvable~\cite{BB-Vortex}, and maximally superintegrable~\cite{Heinzl:2017blq} owing to the existence of other conserved quantities corresponding to non-Poincar\'e symmetries.  An investigation using the BJR-analogue gauge~\cite{Heinzl:2017blq} shows many notable similarities with those of screw-symmetric gravitational waves~\cite{Zhang:2018srn,Zhang:2018gzn}. Here we will reinvestigate the vortex using the analogue of Brinkmann gauge, which will make the connection to the gravitational case completely explicit. 

The key observation is that the vortex of~\cite{BB-Vortex} can be written in terms of a Brinkmann gauge potential with only a single nonzero component,
\be\label{A-VORTEX-DEF}
	A = -H_{ij}(U)X^i X^j \ud U \;,
\ee
with $H_{ij}$ as in (\ref{H-VAL}). Using this we can see that the screw-symmetric gravitational plane wave in Brinkmann coordinates is the classical double copy~\cite{Monteiro:2014cda} of the electromagnetic vortex, as follows. Let $k_\mu$ be the null vector such that $k.x = U$, and define $\phi = -H_{ij}(U)X^i X^j$. Then $\partial_i \partial_i \phi=0$, the vortex potential (\ref{A-VORTEX-DEF}) is $A_\mu = \phi k_\mu$, and the metric (\ref{BRINK}) may be written
\be
	g_{\mu\nu} = \eta_{\mu\nu} + \phi k_\mu k_\nu \;.
\ee
These are the relations for a Kerr-Schild type of double copy~\cite{Monteiro:2014cda,Luna:2015paa,Bahjat-Abbas:2017htu}. This prompts a comparison of particle motion in the gravitational and electromagnetic backgrounds.
%
%
%%%%%%%%%%%%%%%%%%%%%%%%%%%%%%%%%
\subsection{Motion in the electromagnetic vortex}
%%%%%%%%%%%%%%%%%%%%%%%%%%%%%%%%%
%
The classical equations of motion in the vortex, following from the action in (\ref{H-U1}), imply $\ud^2U/\ud\tau^2=0$, so that we may trade proper time $\tau$ for $U$ with $U = 2p_V \tau/m$. We again use a dot for a $U$-derivative. The remaining equations of motion are
\be\begin{split}\label{EOM-U1}
	\ddot{X}^i &= \frac{1}{p_V} H_{ij}(U) X^j \;,  \qquad \ddot{V} = \frac{2}{p_V} H_{ij}(U) \dot{X}^i X^j   \;.
\end{split}
\ee
The equations for the `transverse subsystem' $\{X^1,X^2\}$ decouple~\cite{BB-Vortex}, and can be solved analytically by changing to a rotating frame, such that the equations describe two coupled, but time-independent, oscillators~\cite{BB-Vortex} (see also below). Once the transverse orbit is found, the equation for $V$ can be integrated directly.

Momentum $p_V$ is conserved. There are four more conserved quantities, which take rather complicated, and unrevealing, forms in BJR gauge~\cite{Heinzl:2017blq}, and do not correspond to Poincar\'e symmetries of $F_{\mu\nu}$. Using Brinkmann gauge, on the other hand, makes their origin clear; the conserved quantities are
\be\label{Q14EM}
	Q = f_j(U) p_j + 2 X^j {\dot f}_j(U) p_V \quad \text{where} \quad \ddot{f}_i (U)= \frac{1}{p_V}H_{ij}(U) f_j(U) \;, 
\ee
corresponding precisely to the four symmetries of the gravitational plane wave (\ref{xi-4-grav}); these symmetries are thus shared between the vortex and its classical double copy. Note that $f(U)$ obeys precisely the same equations as the classical transverse orbit~\cite{BB-Vortex,Heinzl:2017blq}. The same is true for the gravitational case, to which we now turn.

%%%%%%%%%%%%%%%%%%%%%%%%%%%%%%
\subsection{Motion in screw-symmetric gravitational waves}
\begin{figure}[t!]
\begin{tabular}{cc}
\begin{minipage}{5cm}
\includegraphics[width=4.5cm]{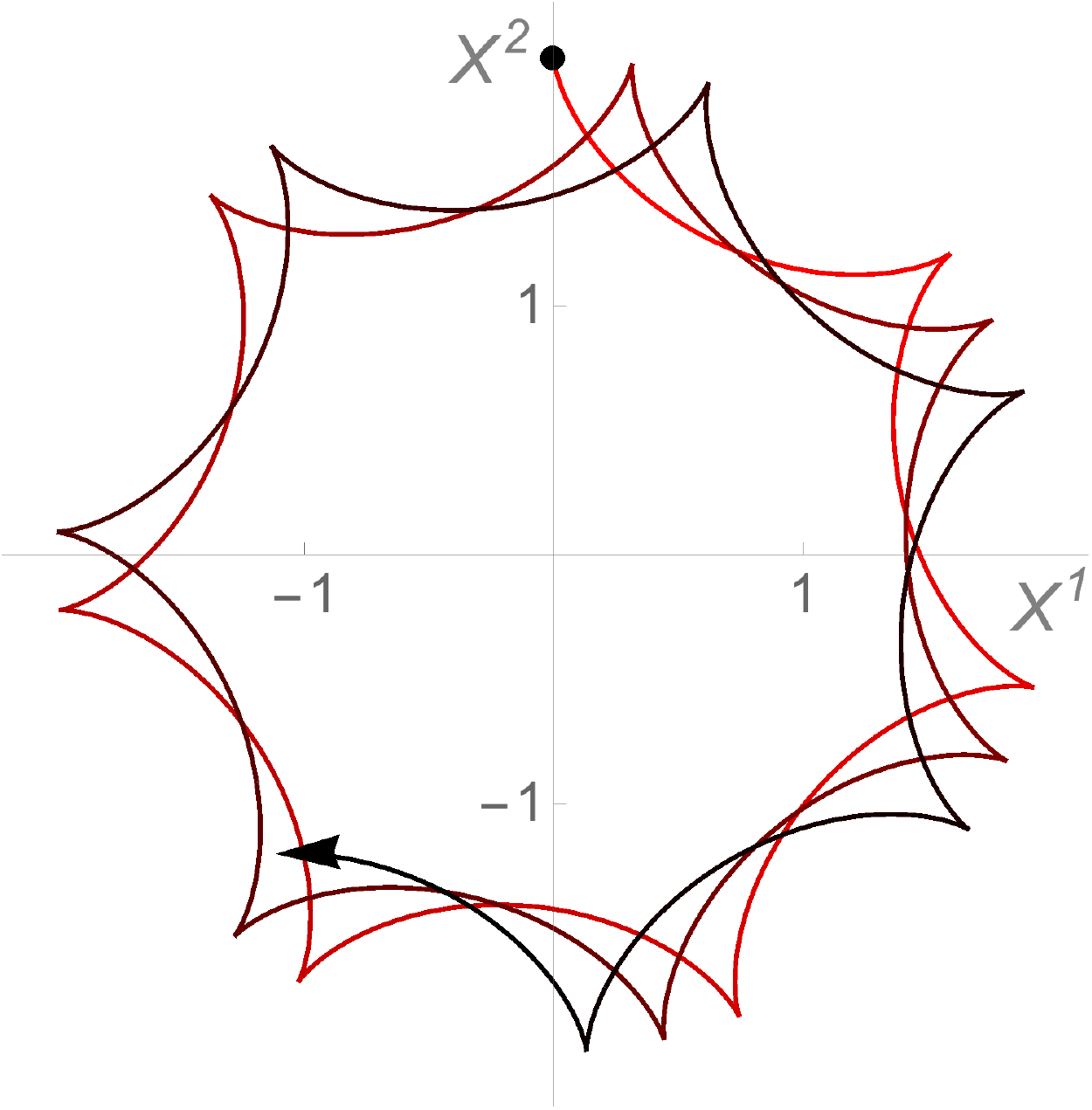}
\begin{align*}
	\{X^1,X^2,V\}\big|_0 =\{0,2,0\} \, \, \, \, \, \, \\
	 \{\dot{X}^1,\dot{X}^2,\dot{V}\}\big|_0 = \{0,0,1/2\}
\end{align*}
\end{minipage}
&
\begin{minipage}{0.45\textwidth}
\raisebox{90pt}{U(1)\; }{\includegraphics[width=\textwidth]{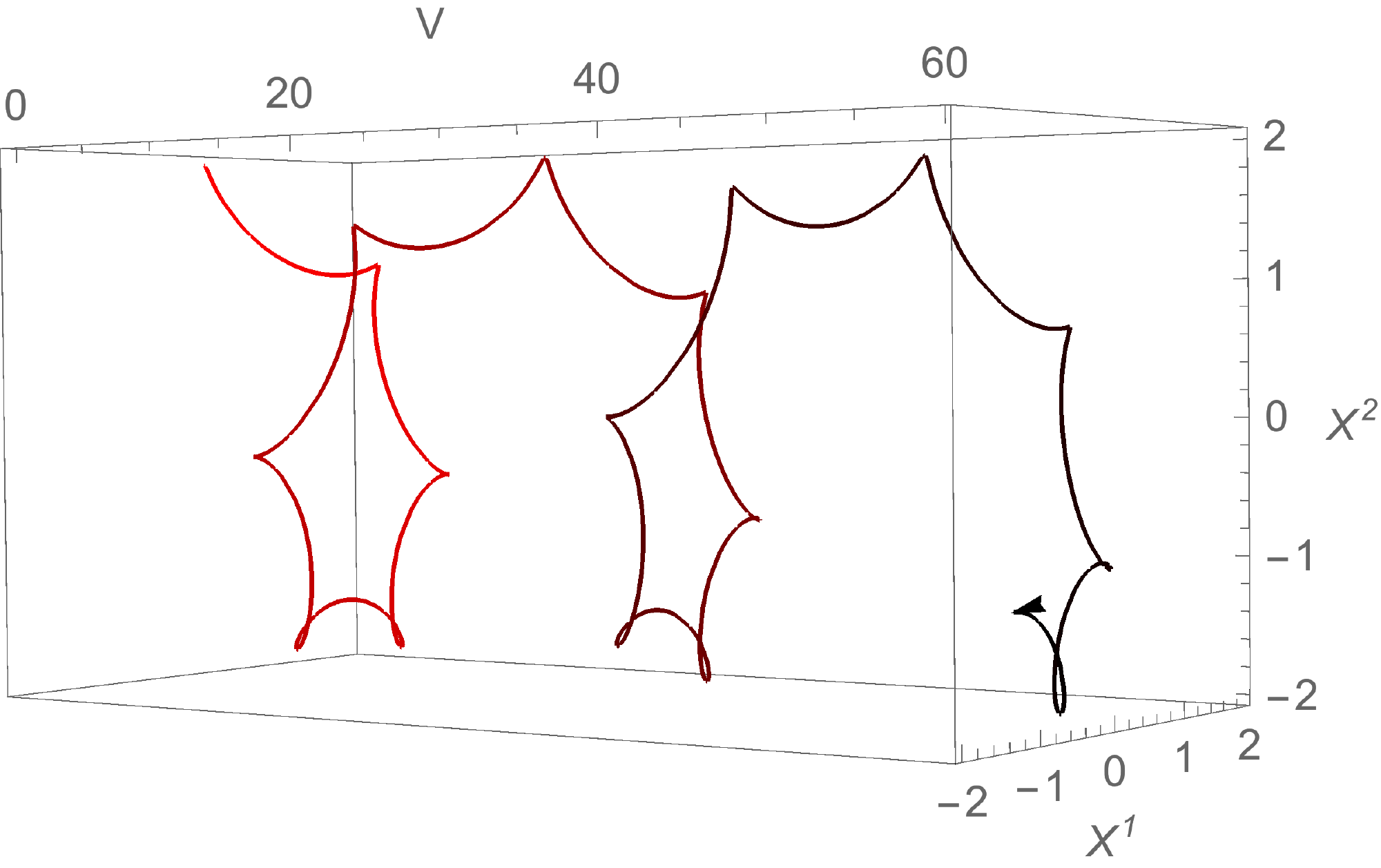}} \\
\raisebox{90pt}{GR\;\;\; }{\includegraphics[width=\textwidth]{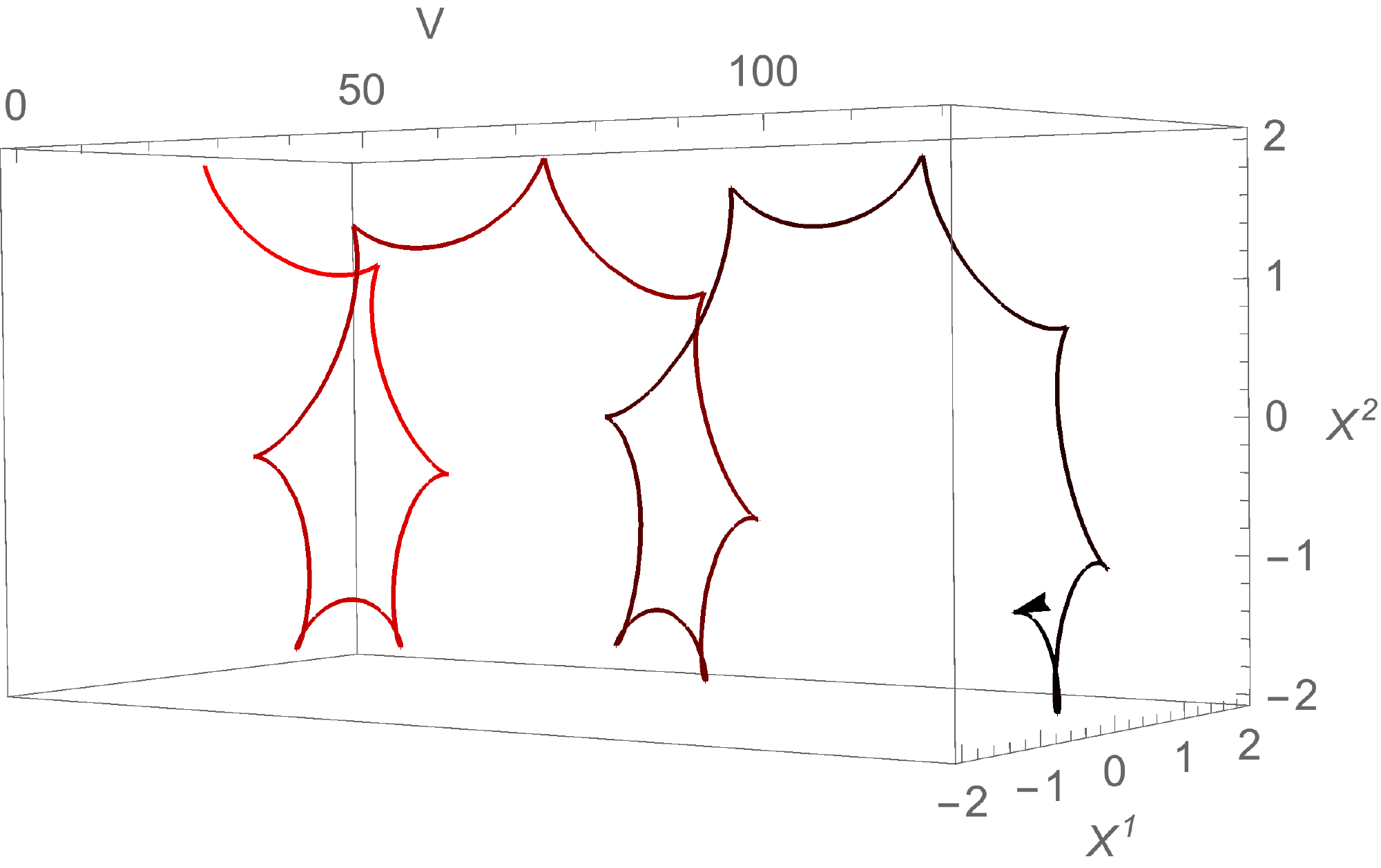}}
\end{minipage}
\end{tabular}
%%%%%%%%%%%%%
\begin{tabular}{cc}
\begin{minipage}{5cm}
\includegraphics[width=4.5cm]{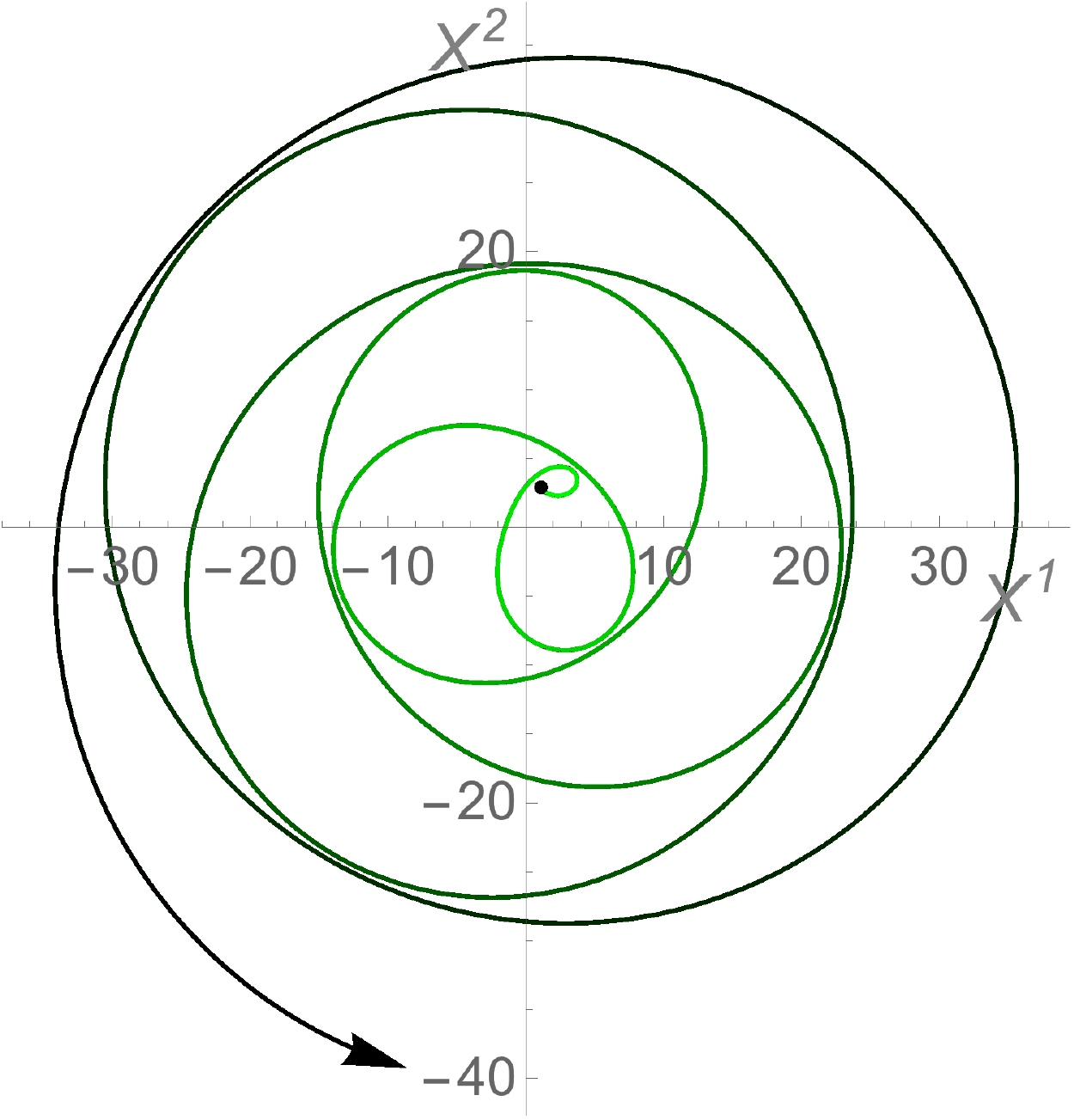}
\begin{align*}
	\{X^1,X^2,V\}\big|_0 =\{1,3,0\} \, \, \, \, \, \, \\
	 \{\dot{X}^1,\dot{X}^2,\dot{V}\}\big|_0 = \{0,0,1/2\}
\end{align*}
\end{minipage}
&
\begin{minipage}{0.45\textwidth}
\raisebox{90pt}{U(1)\; }{\includegraphics[width=\textwidth]{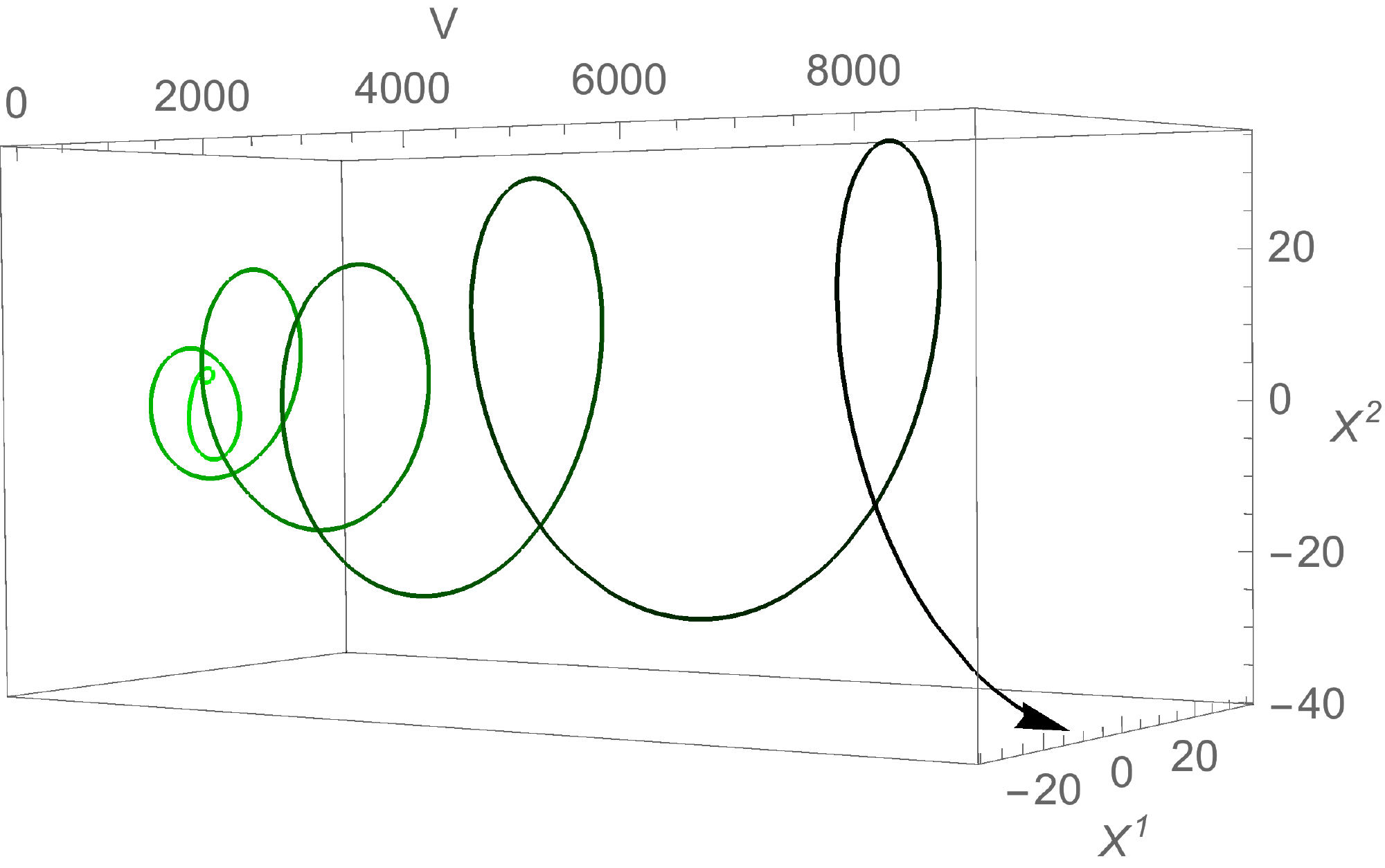}} \\
\raisebox{90pt}{GR\;\;\; }{\includegraphics[width=\textwidth]{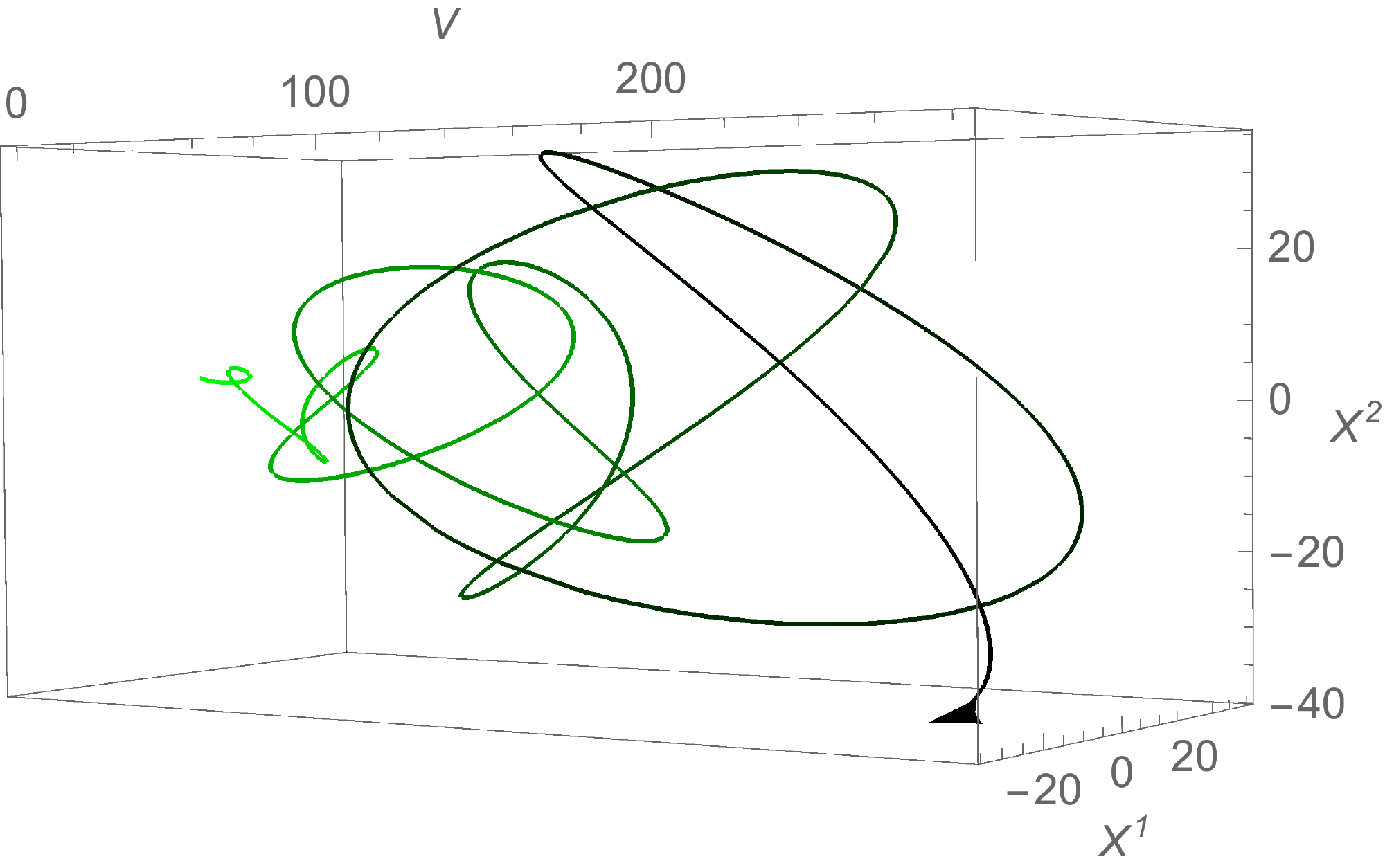}}
\end{minipage}
\end{tabular}
\caption{\label{FIG:1} Comparison of particle motion in an electromagnetic vortex and a screw-symmetric gravitational wave. Motion in the transverse plane $\{X^1,X^2\}$ (left) is identical for matched parameters, here $H^{\text{grav}}_0 = H_0^{U(1)}/(2p_V) = 1/4$. (We use units in which $\omega=1$, other parameters as shown.) The motion is sensitive to initial conditions and may show periodic motion, precession, expanding spirals, or rather involved orbits. Motion in $V$ is in general different in the electromagnetic and gravitational cases, even for matched parameters. It may though be similar, upper panels (but note the scales), or very different, lower panels.}
\end{figure}

%%%%%%%%%%%%%%%%%%%%%%%%%%%%%%
%
We now compare with the motion of a test particle in the screw-symmetric gravitational wave. Using the action in (\ref{H-GR}), we may again parameterise the orbit with $U$.  The remaining equations are, see e.g.~\cite{Zhang:2018srn},
\be\begin{split}\label{EOM-GR}
	\ddot{X}^i &= H_{ij}(U) X^j \;,  \qquad \ddot{V} = 4 H_{ij}(U) \dot{X}^i X^j +  \dot{H}_{ij} X^i X^j \;. 
\end{split}
\ee
The similarity between (\ref{EOM-GR}) and (\ref{EOM-U1}) is apparent; in fact the transverse motion in the two cases is identical if we identify the amplitude $H_{0}$ in (\ref{H-VAL}) in the gravitational case with $H_0/p_V$ in the electromagnetic case. Again the equation for $V$ can be integrated directly, but clearly the orbit in $V$ is different from the electromagnetic case. This is nicely summarised by comparing the corresponding Hamiltonians,
\be
	H_\text{GR} = \frac{p_1^2+p_2^2+m^2}{4 p_V}-p_V H_{ij} X^i X^j \;, \qquad 		H_\text{U(1)} = \frac{p_1^2+p_2^2+m^2}{4 p_V}-H_{ij} X^i X^j  \;, 
\ee	
in which the only difference is the additional factor of $p_V$ in the second term of the gravitational case. This results from the quadratic, rather than linear, coupling of the particle to the background. A comparison of particle motion in the gravitational and electromagnetic fields is shown in Fig.~\ref{FIG:1}.

We return to the question of conserved quantities. Each of the four isometries in (\ref{xi-4-grav}) yields a conserved quantity $Q_1\ldots Q_4$ which, as in the electromagnetic case, is expressed in terms of the classical $X^j$. These quantities have the same form as in the electromagnetic case (\ref{Q14EM}), but where the $f_j$ obey the relation in~(\ref{xi-4-grav}). $Q_5=p_V$ is conserved, and the screw symmetry (\ref{SCREW}) yields the conservation of 
\be
	Q_6 = H + \frac{w}{2}\big( X^1 p_2 - X^2 p_1\big) \;.
\ee
As in the electromagnetic case, this cannot be independent of the other conserved quantities. To show this explicitly we need the classical orbit, but as mentioned above the general solution is rather complicated. Hence we follow the nice observation of~\cite{Zhang:2018srn} that for the particular choice of parameters $H_0=1/4$ in our conventions (in units such that $\omega=1$) the transverse orbits take a simple form, and we present this case.  The method of solution~\cite{Zhang:2011bm,Zhang:2012cr} is the same as in the electromagnetic case~\cite{BB-Vortex,Heinzl:2017blq}. Using the shorthand notation $s_1\equiv \sin(u/2)$, $s_2\equiv \sin(u/\sqrt{2})$, $c_1\equiv \cos(u/2)$, $c_2\equiv \cos(u/\sqrt{2})$ the four independent solutions to the transverse equations are  
%
%%%%%%
\be\begin{split}
\{X^1,X^2\}_1 &= \left\{s_1 \big(u-\sqrt{2} s_2\big)-c_1 \left(c_2-2\right),c_1 \big(\sqrt{2} s_2-u\big)-\left(c_2-2\right) s_1\right\} \;, \\
\{X^1,X^2\}_2 &=\left\{c_1 s_2-\sqrt{2} \left(c_2-1\right) s_1,\sqrt{2} c_1 \left(c_2-1\right)+s_1 s_2\right\} \;, \\
\{X^1,X^2\}_3 &=\left\{-s_1,c_1\right\} \;, \\
\{X^1,X^2\}_4 &=\left\{s_1 \big(u-2 \sqrt{2} s_2\big)-2 c_1 \left(c_2-1\right),-c_1 \big(u-2 \sqrt{2} s_2\big)-2 (c_2-1) s_1\right\} \;.
\end{split}
\ee
With this, we can write down the explicit forms of $Q_1\ldots Q_4$ (and $Q_5$). One may then verify that
\be
	Q_6 = \frac{2 \sqrt{2} Q_2 Q_3 -2 Q_1 Q_4+ m^2+\sum_1^4 Q_j Q_j}{4 Q_5} \;,
\ee
so that the conservation of $Q_6$ is indeed implied by the conservation of the other five.
%%%%%%%%%%%%%%%%%%%%%%%%%%
\section{Conclusions}\label{SECT:CONCS}
%%%%%%%%%%%%%%%%%%%%%%%%%%

Classical particle dynamics in plane waves has long been known to be exactly solvable. In fact this is a rare example of a superintegrable system. The five symmetries common to all plane waves (gravitational or electromagnetic) determine particle motion. They leave no room for additional conserved quantities, and when these are implied by symmetries of the background they cannot be independently conserved. We have confirmed this for both gravitational and electromagnetic waves in the case of screw symmetry.

Taking a general plane wave and imposing invariance under the screw transformation fixes the wave's two arbitrary functions ($H_{ij}$ in the gravitational case). It is possible to write down other examples; imposing invariance under the sum of a translation and a boost, for example, rather than the sum of a translation and a rotation, yields $H_{ij} \sim 1 /(1+ \omega u)^2$.

Such highly symmetric systems make for interesting study. We have observed that particle dynamics in the screw-symmetric gravitational wave has many similarities with that in an electromagnetic vortex, and we have made this connection explicit; the screw-symmetric gravitational plane wave is the classical double copy of the electromagnetic vortex which, we note, is not a plane wave. In fact this yields some intuition for the electromagnetic case; while the conserved quantities in the particle motion were known, we have found here that they have their origin in the isometries of a gravitational wave. \\

\textit{A.I.~thanks Tim Adamo and Tom Heinzl for many useful discussions.}

%%%%%%%%%%%
\section*{References}
%%%%%%%%%%%

\end{document}